\documentclass[aps,prb,twocolumn,superscriptaddress,showpacs]{revtex4-1}

\usepackage{hyperref}
\usepackage[dvips]{graphicx}
\usepackage{amssymb,amsmath,amsfonts}
\usepackage[T1]{fontenc}
\usepackage{float}
\usepackage{color,soul}
\usepackage{booktabs}
\usepackage{color}
\newcommand{\bra}[1]{\langle #1|}
\newcommand{\ket}[1]{|#1\rangle}
\newcommand{\unit}{$(\hbar/e)(\Omega\times $cm$)^{-1}$}
\usepackage{changes}

\begin{document}


\title{Giant spin Hall effect in two-dimensional monochalcogenides}


\author{Jagoda S\l awi\'{n}ska} 
\email{Email: jagoda.slawinska@gmail.com}
\affiliation{Department of Physics, University of North Texas, Denton, TX 76203, USA}

\author{Frank T. Cerasoli}
\affiliation{Department of Physics, University of North Texas, Denton, TX 76203, USA}

\author{Haihang Wang}
\affiliation{Department of Physics, University of North Texas, Denton, TX 76203, USA}

\author{Sara Postorino}
\affiliation{Dipartimento di Fisica, Universit\`a di Roma Tor Vergata, Via della Ricerca Scientifica 1, 00133 Roma, Italy}

\author{Andrew Supka}
\affiliation{Department of Physics and Science of Advanced Materials Program, Central Michigan University, Mount Pleasant, MI 48859, USA}
\affiliation{Center for Materials Genomics, Duke University, Durham, NC 27708, USA}

\author{Stefano Curtarolo}
\affiliation{Center for Materials Genomics, Duke University, Durham, NC 27708, USA}
\affiliation{Materials Science, Electrical Engineering, Physics and Chemistry, Duke University, Durham, NC 27708, USA}

\author{Marco Fornari}
\affiliation{Department of Physics and Science of Advanced Materials Program, Central Michigan University, Mount Pleasant, MI 48859, USA}
\affiliation{Center for Materials Genomics, Duke University, Durham, NC 27708, USA}

\author{Marco \surname{Buongiorno Nardelli}} 
\email{Email: mbn@unt.edu}
\affiliation{Department of Physics, University of North Texas, Denton, TX 76203, USA}
\affiliation{Center for Materials Genomics, Duke University, Durham, NC 27708, USA}

\vspace{.15in}

\date{\today}

\begin{abstract}
One of the most exciting properties of two dimensional materials is their sensitivity to external tuning of the electronic properties, for example via electric field or strain. Recently discovered analogues of phosphorene, group-IV monochalcogenides (MX with M = Ge, Sn and X = S, Se, Te), display several interesting phenomena intimately related to the in-plane strain, such as giant piezoelectricity and multiferroicity, which combine ferroelastic and ferroelectric properties. Here, using calculations from first principles, we reveal for the first time giant intrinsic spin Hall conductivities (SHC) in these materials. In particular, we show that the SHC resonances can be easily tuned by combination of strain and doping and, in some cases, strain can be used to induce semiconductor to metal transition that makes a giant spin Hall effect possible even in absence of doping. Our results indicate a new route for the design of highly tunable spintronics devices based on two-dimensional materials. 
\end{abstract}

\maketitle

\section{Introduction}
 The spin Hall effect (SHE) is a phenomenon emerging from spin-orbit coupling (SOC) in which an electric current or external electric field can induce a transverse spin current resulting in spin accumulation at opposite sample boundaries.\cite{she1, kato, she3, sinova_she} The charge/spin conversion without the need of applied magnetic fields makes the SHE an essential tool for spin manipulation in any spintronics devices\cite{sinova, fujiwara} and the subject of intensive theoretical and experimental research. The intrinsic SHE in crystals was predicted and observed experimentally in a variety of materials, ranging from doped semiconductors (GaAs)\cite{kato} to elemental metals with strong SOC, such as platinum, tantalum, palladium and tungsten.\cite{pt, simple_she, au_pd, pt_w, tantalum, tungsten, experimental} It has been also investigated in metallic and semimetallic thin films\cite{sriro3}  where SHC can be enhanced with respect to the corresponding bulk phase. Studies related to SHE in two dimensional (2D) materials are limited to only few works focused on transition metal dichalcogenides\cite{tmd} and simple material models.\cite{model1}


In this paper, a giant intrinsic SHE tunable by combination of strain and doping is predicted for the first time in monolayer group-IV monochalcogenides MX with M = Ge, Sn and X = S, Se, Te, often referred to as analogues of phosphorene due to their structural similarity.\cite{gomes, enhanced_piezo, strain} The bulk parent compound has the orthorhombic crystal structure of black phosphorus ($Pnma$) and consists of weakly bonded van der Waals layers, making an exfoliation process a viable route to produce atomically thin films or single layer crystals; indeed, some materials from this family have been already synthesized experimentally. \cite{exp1, exp2, exp3, science_snte, parkin} As most 2D materials, group-IV monochalcogenides exhibit several extraordinary mechanical, electronic and optical properties. These include high flexibility, large thermal conductivity, giant piezoelectricity \cite{apl}, multiferroicity,\cite{multiferroics_2dmater, multiferroics_nano} superior optical absorbance,\cite{optics, water, xu} and even valley Hall effect,\cite{valley} making them promising candidates to use in multifunctional devices. Finally, the strong SOC suggests their high potential for spintronics. 

\begin{figure}[h!]
    \includegraphics[width=\columnwidth]{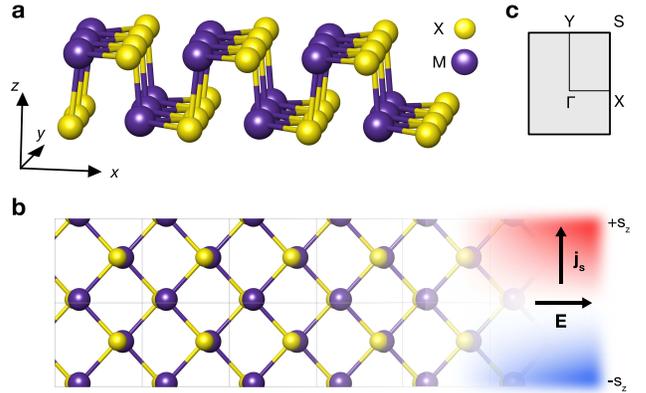}
    \caption{\label{struct}
    Structure of 2D group IV monochalcogenides. Panels (a) and (b) show side and top views, respectively, while the scheme of Brillouin zone is displayed in (c). Right-hand side of panel (b) additionally illustrates an example of geometry setup for spin Hall effect possible to realize in doped monolayers. We note that only $\sigma^{z}_{xy}$ and $\sigma^{z}_{yx}$ components of SHC tensor are different from zero.
    }
\end{figure}

The group-IV monochalcogenides possess wide band gaps, which precludes existence of non-negligible spin Hall conductivity at intrinsic chemical potential, similarly to conventional bulk semiconductors where significant electron or hole doping is needed in order to achieve a measurable spin Hall effect. Although our calculations show that a giant SHC could be reached with $p$- or $n$-type doping of $n_{h/e}=1\times10^{14}$ e/cm$^{2}$, which is an order of magnitude lower than in case of transition metal dichalcogenides,\cite{tmd} such values may still be difficult to reach experimentally. Here, we propose an alternative route to realize SHE in these materials. We demonstrate that compressive or tensile strain along any axis not only can tune the position of the SHC resonances, but can also induce semiconductor to metal transitions that make a giant spin Hall effect possible even in absence of doping. As such, different phases of SHE can be switched externally via strain allowing direct engineering of spintronics functionalities in these materials. 

\section{Methods}
Our noncollinear DFT calculations were performed using the \textsc{Quantum Espresso} code\cite{qe,qe1} interfaced with the \textsc{AFLOW$\pi$} and \textsc{PAOFLOW} computational infrastructures.\cite{aflowpi,paoflow} We used the generalized gradient approximation (GGA) in the parametrization of Perdew, Burke, and Ernzerhof (PBE)\cite{pbe} and, to further improve the description of the electronic properties, a novel pseudo-hybrid Hubbard self-consistent approach ACBN0.\cite{acbn0} The ion-electron interaction was treated with the projector augmented-wave fully-relativistic pseudopotentials\cite{kresse-joubert} from the pslibrary database\cite{pslibrary} while the wavefunctions were expanded in a plane-wave basis of 50 Ry (500 Ry for the charge density). The Brillouin zone sampling at DFT level was performed following the Monkhorst-Pack scheme using a $24\times24\times2$ k-points grid, further increased to $140\times140\times2$ with PAOFLOW's Fourier interpolation method to accurately integrate spin Berry curvatures. 

The intrinsic spin Hall conductivities were calculated using the \textsc{PAOFLOW} code\cite{paoflow} following the linear-response Kubo-like formula:\cite{gradhand, kubo, macdonald_fe}

\begin{eqnarray}
    \sigma^{s}_{ij} = \frac{e^2}{\hbar}\sum_{\vec{k}}\sum_{\vec{n}}f_n(\vec{k})\Omega^{s}_{n,ij}(\vec{k})  \nonumber \\ 
    \Omega^{s}_{n,ij} (\vec{k}) = \sum_{m\neq n}\frac{2\textrm{Im}\bra{\psi_{n, \vec{k}}}j^{s}_{i}\ket{\psi_{m, \vec{k}}}\bra{\psi_{m, \vec{k}}}v_{j}\ket{\psi_{n, \vec{k}}}}{(E_n-E_m)^2}\nonumber
\end{eqnarray}
where $\vec{j} = \{s, \vec{v}\}$ is the spin current operator with $s = \frac{\hbar}{2}(\beta,\Sigma : 4\times4 \textrm{ Dirac matrices})$ and $f_n(\vec{k})$ is the Fermi distribution function for the band $n$ at $\vec{k}$. We note that, in contrast to most of reported calculations of SHC based on the above formula,\cite{macdonald_fe, simple_she, tmd} we do not add any infinitesimal term $\delta$ in the denominator to avoid singularities if the bands are degenerate. We have evidence (see Fig. S1 in Supplementary Material (SM)) that using a finite $\delta$ in Kubo's formula leads to the unphysical behavior of non-zero values of SHC within the semiconductor's gap whose origin was unclear so far. Using perturbation theory for degenerate states to avoid numerical singularities ensures that $\sigma^{s}_{ij}$ always vanishes at the Fermi level. 

Figure \ref{struct} shows the 2D non-centrosymmetric unit cell of the phosphorene-like phase used in the calculations for all six compounds. It contains four atoms arranged in two buckled layers resembling a monolayer of black phosphorus with a mirror symmetry axis along $x$, representing one of the four ground states of the system.\cite{multiferroics_2dmater, multiferroics_nano, new_monochal} The lattice constants and ionic positions were fully relaxed without including SOC whose influence on forces is known to be negligible. The electronic structure was then recalculated with SOC self-consistently. The vacuum region of ~20 \AA\, was set to prevent any interaction between spurious replicas of the slab. The configurations with the relaxed lattice constants were used as a starting point for the simulations of strained structures; we considered strains varying between -10\% to 10\% along $x$ and $y$ axis simultaneously, in each case relaxing the positions of the atoms. We analyzed, in total, 726 different structures. Further details of the calculations as well as additional results, including lattice constants, values of band gaps and convergence tests for SHC are reported in the SM (Table S1, Fig.S1).

\section{Spin Hall effect in unstrained monolayers}
\begin{figure*}[ht!]
    \includegraphics[width=\textwidth]{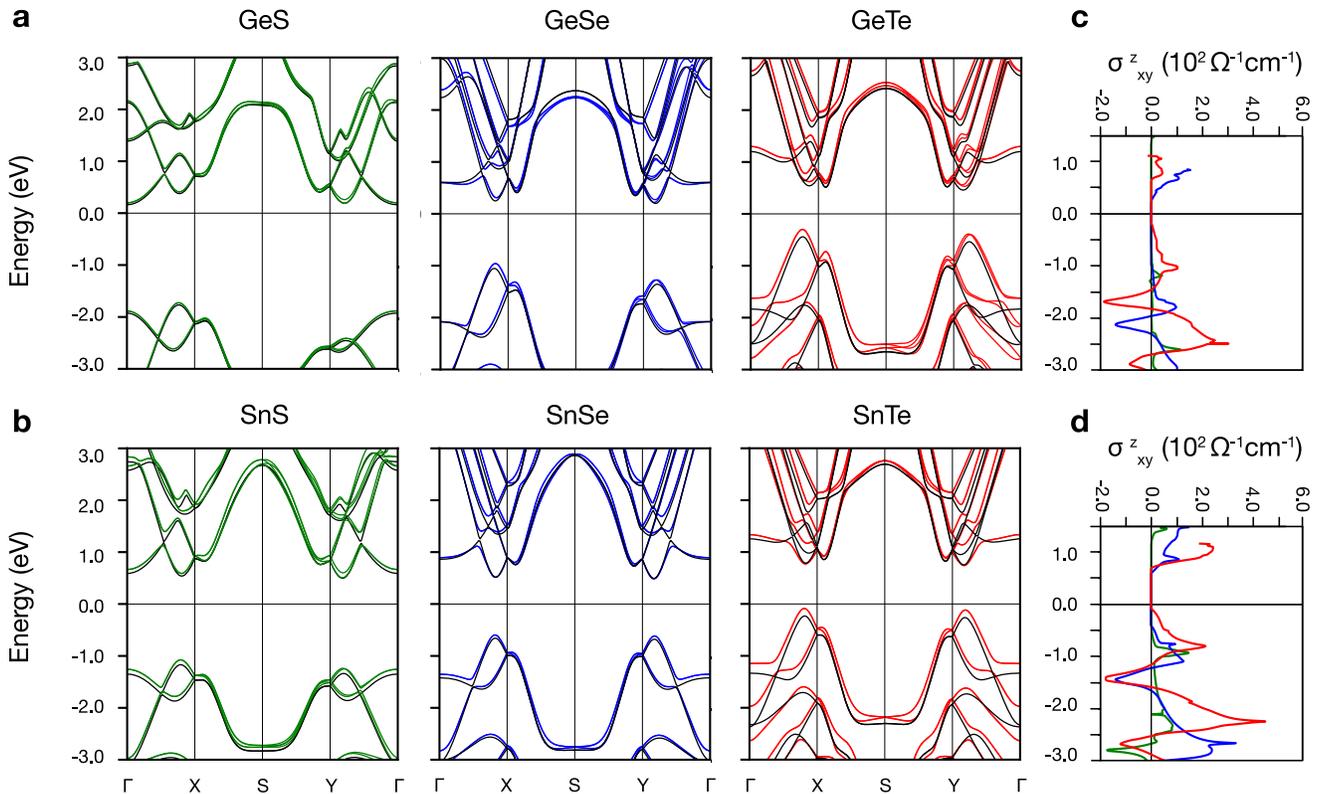}
    \caption{\label{bands}
    Relativistic electronic structures of group IV monochalcogenides GeX (a) and SnX (b), X = S, Se, Te represented as green, blue and red lines, respectively. The corresponding scalar-relativistic band structures are superimposed (black lines). (c-d) Spin Hall conductivities $\sigma^{z}_{xy}$ calculated as a function of chemical potential for compounds in panels (a-b) employing the same color scheme. 
    }
\end{figure*}

Let us first consider unstrained structures. Figures \ref{bands} (a-b) summarize the relativistic band structures of sulfides, selenides and tellurides (green, blue and red lines, respectively), while panels (c-d) display their corresponding spin Hall conductivity as a function of chemical potential. Due to the reduced symmetry of two-dimensional structures, the only non-vanishing independent component is $\sigma_{xy}^{z}=-\sigma_{yx}^{z}$. Our band structures and the values of the band gaps calculated with the ACBN0 functional are in good agreement with existing simulations that use hybrid functionals and with available experimental data (see SM, Table S1). Comparison of the scalar-relativistic band structures (plotted as black lines in panels (a-b)) with the fully relativistic ones, clearly indicates a strong impact of SOC, which induces several anti-crossings and splittings of the bands. These effects are moderate in selenides and  most pronounced in tellurides given the heaviness of Te atoms. Although it is quite difficult to attribute particular features of the relativistic band structures to the specific peaks (resonances) in $\sigma_{xy}^{z} (E)$, one can easily observe that severe SOC-induced modifications in the electronic structure of GeTe and SnTe result in giant values of spin Hall conductivity, $\approx$ 300 and 500 \unit, respectively, for higher binding energies. We note that even the resonances closest to the Fermi level ($E_F$) still achieve values as large as 200 \unit. The selenides exhibit slightly lower ($\sim$100) magnitudes of SHC and the sulfides do not seem to display any spin Hall effect at all. 

As we have mentioned above, similarly to other semiconductors, either $p-$type or $n-$type doping is needed to reach the SHC resonances. In Table I, we list the values of the SHC peaks and the corresponding doping levels expressed as a Fermi level shift and number of electrons per surface unit, for six compounds reported in Fig. \ref{bands}. In general, the doping concentrations are of the order of $n_{h/e}=10^{14}$ e/\textrm{cm}$^{2}$, an order of magnitude lower than in the case of recently studied transition metal dichalcogenides\cite{tmd} but still beyond the typical values achieved in experiments ($\sim 10^{12}-10^{13}\,e/\textrm{cm}^{2}$). However, for the compounds with highest SHC peaks, the spin Hall effect could be very large even at lower doping: for instance, in SnTe the SHC reaches 100 \unit\ for doping of $\sim 10^{13}\,e/\textrm{cm}^{2}$. We also note that the estimated values of doping listed in Table I are simply derived from the density of states following similar analysis in previous theoretical works dealing with SHE in semiconductors\cite{tmd} and the carrier concentrations in real samples might be different. Therefore, we believe that the intrinsic spin Hall effect could be achieved experimentally even in the unstrained structures. 

\setlength{\tabcolsep}{0.8em} 
\begin{table}
\caption{Resonance values of spin Hall conductivity and the corresponding values of doping expressed as Fermi level shift $\Delta E$ and hole/electron concentrations per surface unit $n_{h/e}$. Spin Hall conductivity $\sigma^{z}_{xy}$ is expressed in \unit, $\Delta E$ in eV, and $n_{h/e}$ in $10^{14}$ $e$/cm$^{2}.$ 
}

\begin{tabular}[t]{lcccccc}
\hline
&GeS&GeSe&GeTe&SnS&SnSe&SnTe\\
\hline
$\sigma^{z}_{xy}$&43    &98    &103   &148   &128   &215\\
$E_1$            &-1.18 &-1.86 &-1.00 &-0.93 &-1.09 &-0.79\\
$n_{h}$          &7.12  &5.61  &6.24  &7.34  &4.70  &4.78\\
\hline
$\sigma^{z}_{xy}$&-    &92    &45    &61   &110   &245\\
$E_2$            &-    &0.62  &0.79  &1.45 &0.86  &1.09\\
$n_{e}$          &-    &4.25  &0.74  &1.70 &0.74  &2.56\\
\hline
\hline
\end{tabular}
\end{table}

\section{Spin Hall effect in strained monochalcogenides}
As a next step, we have explored the possibility of tuning the electronic properties and spin Hall conductivity via external strain. The band gap ($E_{g}$) manipulation in group-IV sulfides and selenides was previously reported in Ref. \onlinecite{strain}, where compressive strains along either $x$ or $y$ axis were found to strongly reduce the band gap, and for larger strains could induce a semiconductor to semimetal transition. In the present study, we consider all possible strain configurations varying between $-10\%$ to $+10\%$ with the step of 2$\%$, thus 121 different configurations for each compound. Below, we will discuss only the results for SnTe which displays the highest potential for spintronics; the complete set of results for all considered group-IV monochalcogenides is reported in the SM (Figs. S3-S7). 

\begin{figure*}[ht!]
    \includegraphics[width=\textwidth]{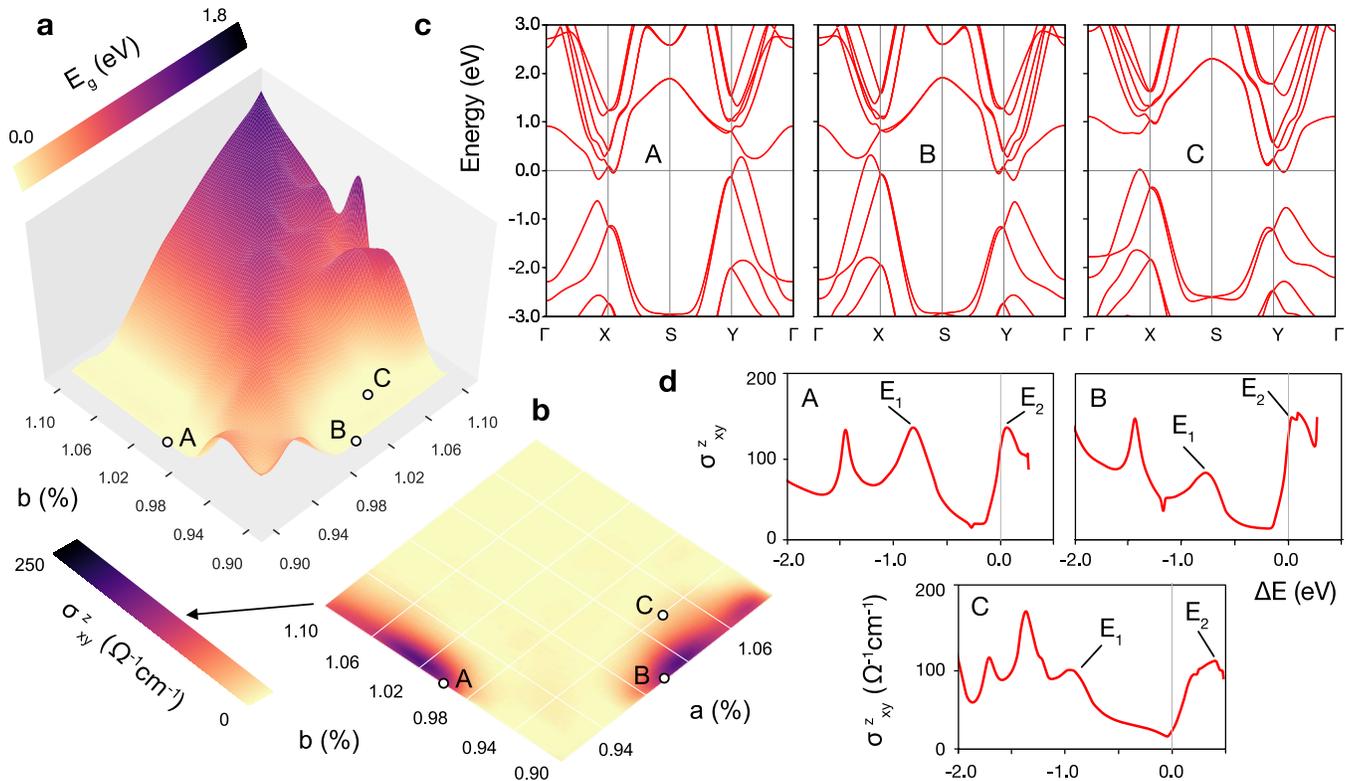}
    \caption{\label{shc_metal}
Electronic properties and spin Hall conductivities of SnTe as a function of strains. (a) Band gap $E_{g}$ vs strains of $a$ (along $x$ axis) and $b$ (along $y$ axis) displayed as a 3D surface in lattice constants space. The corresponding legend is shown in the upper left corner. (b) Heat-map of spin Hall conductivities calculated at zero chemical potential for each strain configuration. The legend is displayed in the bottom left corner. The reversed contrast of maps (a) and (b) clearly reflects their physical meaning and the fact that $\sigma^{z}_{xy}$ can have a finite value only for $E_{g} = 0$. Since uniaxial strain is often considered in experimental realizations, we have also introduced complementary plots summarizing the influence of uniaxial strains along $x$ and $y$ on the band gap and the SHC (see SM, Fig. S2). (c) Band structures of SnTe calculated for selected points in strain space marked in (a-b), A: $a$ = 0.9$a_{0}$ , $b$ = 1.0$b_{0}$, B: $a$ = 1.0$a_{0}$ , $b$ = 0.9$b_{0}$, C: $a$ = 1.04$a_{0}$ , $b$ = 0.94$b_{0}$, where $a_0$ and $b_0$ denote original (unstrained) lattice constants. (d) Corresponding $\sigma^{z}_{xy}$ calculated as a function of chemical potential. Labels $E_1$ and $E_2$ at each curve denote resonances of SHC closest to the Fermi level. 
    }
\end{figure*}

\begin{figure*}[ht!]
    \includegraphics[width=\textwidth]{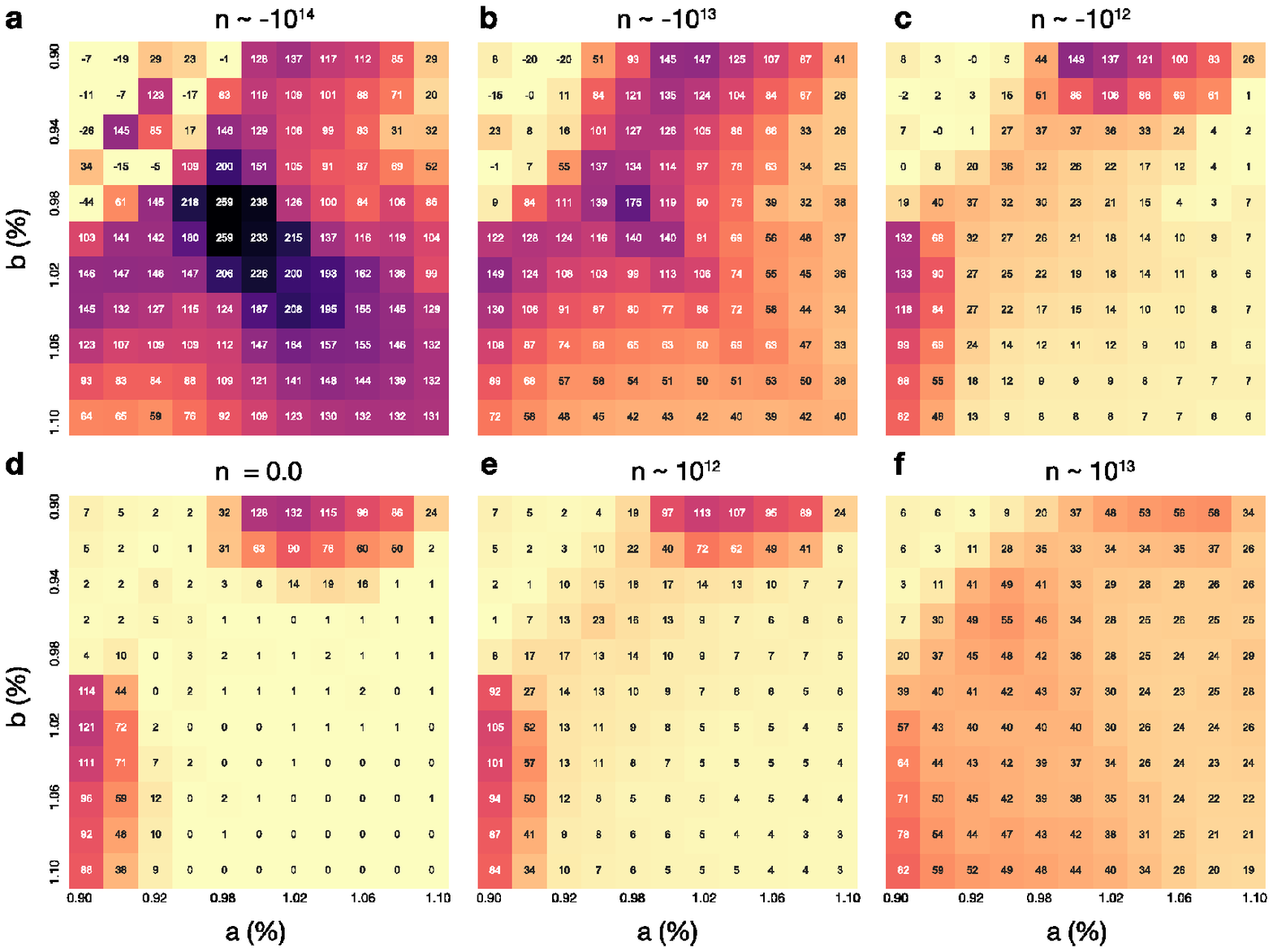}
    \caption{\label{doping}
Spin Hall conductivity of SnTe as a function of strain and doping. (a) $\sigma^{z}_{xy}$ averaged over the values of chemical potential which correspond to doping of order $\sim 10^{14} e/$cm$^2$ displayed as a function of strains along $x$ and $y$. (b-c) Same as (a) for $n_{e} = 10^{13} e/$cm$^2$ and $n_{e} = 10^{12} e/$cm$^2$, respectively (d) $\sigma^{z}_{xy}$ at intrinsic chemical potential, numerically identical to the map in Fig.\ref{shc_metal}(b). Non-zero values for the regions of $E_{g}\neq 0$ are related to the numerical accuracy of calculations. (e)-(f) Same as (c-b) for $p$-type doping. Color scheme same as in Fig.\ref{shc_metal}(b) in all the maps.
} 
\end{figure*}

\subsection{SHE in the metallic phase}
Figure \ref{shc_metal} summarizes the electronic properties and spin Hall conductivity in SnTe for each considered strain configuration. The band gap landscape displayed in panel (a) shows several interesting features: (i) compressive (tensile) strain always leads to decrease (increase) in $E_{g}$, (ii) the lowest values of $E_{g}$ are achieved when a compressive strain along {\it only one} of two axis is applied, (iii) the combinations of tensile and compressive strain can also lead to a decrease in $E_{g}$, larger than in case of compressive strains applied along both axis. We have observed similar behavior in all compounds; in selenides the band gaps are in general wider, thus larger strains are required to enable the semiconductor to semimetal transition. In sulfides, in contrast, a metallic phase cannot be achieved, in agreement with conclusions of Ref. \onlinecite{strain}. 

The corresponding spin Hall conductivities calculated at intrinsic chemical potential plotted in panel (b) clearly reflect the profile in (a), that is, as long as the material is semiconducting, it cannot exhibit any spin Hall effect. Within the regions of $E_{g} = 0$, the values of $\sigma^{z}_{xy}$ vary because each of these strain configurations induces different modifications in the electronic structure. However, we are still able to draw general conclusions regarding the impact of the strain based on the analysis of few selected configurations A, B, C displayed in Fig. \ref{shc_metal} (c) which embody rather huge modifications of both electronic and spin properties. First of all, it is clear that the same strain applied along axis $x$ and $y$ (configurations A and B) affect the dispersion of the bands in an asymmetric way. The strain along $x$ (panel A), in general, brings upwards the occupied bands along the $S-Y-\Gamma$ paths and downwards the unoccupied spectrum along $\Gamma-X-S$ resulting in $p-$type pockets near Y and $n$-type pockets near X, while $y$-strain (panel B) is found to cause opposite shifts. The band structure of configuration C confirms that such tendency is more general; this lower strain configuration is structurally similar to B and indeed its electronic properties are very similar to the latter. 

The spin Hall conductivities of structures A, B, C shown in Fig. \ref{shc_metal} (d) significantly differ from those of the unstrained structure in Fig.\ref{bands}(d), which is not surprising, given the substantial modification of electronic structure at the Fermi level. In order to facilitate a systematic analysis, we have introduced the labels $E_{1}$ and $E_{2}$ corresponding to the two SHC resonances below and above $E_F$, and we followed their behavior due to the electronic structure changes induced by the strain (the positions of $E_{1}$ and $E_{2}$ without strains are reported in Table I).

It is evident that the $\sigma^{z}_{xy}$ can be giant without any doping (structures A, B), which we attribute to the presence of resonance $E_{2}$ much closer to the Fermi level than in the unstrained monolayer ($E_{F}$ is located on the slope in both A and B configurations). Statistical analysis of these two parameters for all calculated structures confirms that indeed the change in the band gap is mainly correlated with the resonance $E_{2}$ (calculated Pearson's correlation between $E_{g}$ and $E_{2}$ is around 80$\%$), while the position of $E_{1}$ hardly depends on the band gap. This means that strain is more likely to shift/modify unoccupied bands which will also determine the SHC. Finally, the $\sigma^{z}_{xy}$ in the moderately strained configuration C also exhibits a finite value at the Fermi level, but since the peak $E_{2}$ is not so close to $E_F$, the spin Hall effect is weaker. 

\subsection{Tuning of SHC via strain and doping}
The experimental realization of SHC, its tuning and switching on/off, is likely to require a combination of doping and strain. In order to quantitatively estimate the effect of both we have calculated the values of $\sigma^{z}_{xy}$ averaged over the range of chemical potential that correspond to a given electron/hole concentration for every configuration of strain. The results for SnTe are shown in Fig. \ref{doping}. In accordance with Table I, $n$-type doping of $\sim 10^{14} ~e/\textrm{cm}^{2}$ (a) guarantees giant values of SHC for unstrained and weakly strained structures. Surprisingly, a compressive strain (a) leads to the reduction of SHC rather than to its increase, which is related to the oscillating character of the intrinsic spin Hall conductivity as a function of chemical potential. In this case, the combination of doping and compressive strain shifts the Fermi level excessively, well beyond the resonance peak of SHC. Moreover, it is clear that a biaxial strain could be used to switch on/off the SHC. We can observe similar behavior also for $n$-type doping of $\sim 10^{13}~ e/\textrm{cm}^{2}$ (b), in such case the values of SHC are lower, but can still be considered large even for unstrained/weakly strained structures. For $n_{e} = \sim 10^{12}~ e/\textrm{cm}^{2}$ (c) the finite but small values of SHC can be increased by uniaxial compressive strain; this low doping configuration resembles the properties of the undoped structures (d) and weakly $p-$type doped configuration (e). Further increase in $p-$type doping to $n_{h} = \sim 10^{13}~ e/\textrm{cm}^{2}$ (f) offers a possibility of even broader modulation of SHC by strain; while biaxial compressive strain can result in switching off the SHC, unaxial strain leads to its increase. Overall, however, the values of SHC are lower here than in case of electron doping. Thus, as anticipated in the previous section, the manipulation of SHE seems to be more feasible in $n$-type doped systems. Finally, among the other monochalcogenides, GeTe reveals interesting properties for spintronics, while the sulfides and selenides either do not possess large SHC at all or it does not exhibit sufficiently high tunability (see SM, Figs. S4-S7). 

\section{Summary and conclusions}
In summary, we have reported for the first time the emergence of a giant spin Hall effect in group IV monochalcogenides, which can be switched on/off and modulated either by doping or uniaxial compressive strain. The most interesting candidate for spintronics is SnTe. We have predicted that the SHE in this compound can be very strong. Moreover, the monolayer and multilayer samples have been recently synthesized and they reveal high potential for technology.\cite{science_snte, silvia_snte, parkin} While our work is limited to monolayers structurally similar to black phosphorus, it is worthwhile to mention that the multilayer stacks can exhibit more intriguing spin-orbit related properties and with even broader possibilities of tuning. Finally, SHE has been achieved in bulk $\beta-$SnTe which suggests that its successful realization in 2D phase is very probable.\cite{ohya} Despite the values of doping/strain required to reach/tune the giant SHC might seem large, the actual 2D character of these compounds can greatly help to overcome these difficulties. For example, different stacking order or thickness of the multilayer might reduce the required doping or the Fermi level could be additionally shifted by any charge originating from a substrate,\cite{carbon} or any additional strain could result from lattice constant matching at the interface. 

Finally, we emphasize that 2D spintronics is not just a hypothesis; the recent discovery of 2D ferromagnetism\cite{cri, crgete} brings it much closer to realization and new candidate materials will be needed. The giant SHC of monochalcogenides combined with low charge conductivity even in metallic phase suggest they might be useful in versatile spintronics applications, such as spin detectors,\cite{fujiwara} and even more likely in novel multifunctional devices. We believe that the properties unveiled in this paper clearly show the high potential of 2D phosphorene analogues for spintronics, and will trigger the interest in the experimental realization of spin Hall effect in these materials. 

{\it Acknowledgments.} We would like to thank Ilaria Siloi and Priya Gopal, for useful discussions. The members of the AFLOW Consortium  (http://www.aflow.org)
acknowledge support  by DOD-ONR (N00014-13-1-0635, N00014-11-1-0136,
N00014-15-1-2863). MBN and FTC acknowledge partial support from Clarkson Aerospace Corporation. The authors also acknowledge Duke University --- Center for Materials Genomics --- and the CRAY corporation for
computational support. Finally, we are grateful to the High Performance Computing Center at the University of North Texas and the Texas Advanced Computing Center at the University of Texas, Austin.

\end{document}